# The Effect of Spin on the Flight of a Baseball

Alan M. Nathan[1], Joe Hopkins, Lance Chong, and Hank Kaczmarski
University of Illinois at Urbana-Champaign

[1]a-nathan@uiuc.edu

**Abstract.** New measurements are presented of the lift on a spinning baseball. The experiment utilizes a pitching machine to project the baseball horizontally; a high-speed motion capture system to measure the initial velocity and angular velocity and to track the trajectory over ~5 m of flight; and a ruler to measure the total distance traversed by the ball. Speeds in the range v=50-110 mph and spin rates (topspin or backspin) $\omega$=1500-4500 rpm were utilized, corresponding to Reynold's numbers Re=(1.1-2.4) x $10^5$ and spin parameters $R\omega/v$=0.1-0.6. Standard least-squares fitting procedures were used to extract the initial parameters of the trajectory and to determine the lift and drag coefficients. Comparison is made with previous measurements or parametrizations of the lift and drag coefficients, and implications for the effect of spin on the flight of a long fly ball are discussed.

## 1 Introduction to the Problem

In a recent paper, Sawicki et al. (Sawicki, Hubbard, and Stronge 2003) report a study of the optimum bat-swing parameters that produce the maximum range on a batted baseball. Using a model for the ball-bat collision and recent experimental data for the lift and drag coefficients, they tracked the ball from collision to landing. For given initial parameters of the pitch (speed, angle, and spin), the bat swing angle and undercut distance were varied to maximize the range. The study found the surprising result that an optimally hit curveball travels some 12 ft. farther than an optimally hit fastball, despite the higher pitched-ball speed of the fastball. The essential physics underlying this result has to do the with the aerodynamic lift force on a baseball projected with backspin. In general, a baseball will travel farther if it projected with backspin. It will also travel farther if is projected with higher speed. In general a fastball will be hit with a higher speed. However, a curveball will be hit with larger backspin. The reason is that a curveball is incident with topspin and hence is already spinning in the right direction to exit with backspin. A fastball is incident with backspin so the spin direction needs to reverse to exit with backspin. It then becomes a quantitative question as to which effect wins: the higher speed of the fastball or the larger backspin of the curveball. According to Sawicki et al., hereafter referred to as SHS, the latter effect wins and the curveball travels farther.

The conclusion of SHS depends critically on the size of the lift force on a spinning baseball. SHS used a particular model for the lift based largely on experimen-tal data that will be reviewed in the next section. That model and the conclusions that follow have been criticized by Adair (Adair 2005), who claims that SHS grossly overestimate the effect of spin on the flight of a baseball. The goal of the present paper is to resolve the disagreement between SHS and Adair, hereafter referred to as RKA, by performing new measurements of the effect of spin on the flight of a baseball. We first present a more detailed discussion of the differences in Section 2. Our experiment, including the data reduction and analysis, is described in Section 3. Our results are compared to previous determinations of the lift and drag in Section 4 and the implications for the flight of a baseball are discussed. We conclude with a summary in Section 5.

## 2 Previous Determinations of Lift

When a spinning baseball travels through the atmosphere, it experiences the force of gravity in addition to the aerodynamic forces of drag and lift, $F_D$ and $F_L$. Conventionally these forces are parametrized as

$$F_D = ½C_D\,\rho A v^2 \qquad F_L = ½C_L\,\rho A v^2 \tag{1}$$

where A is the cross sectional area of the ball, v is the speed, $\rho$ is the air density, and $C_D$ and $C_L$ are the drag and lift coefficients. SHS and RKA have distinctly different prescriptions for $C_L$. SHS uses the parametrization

$$\text{SHS:} \quad C_L = 1.5S \quad (S<0.1)$$

$$= 0.09 + 0.6S \quad (S>0.1) \tag{2}$$

where $S=R\omega/v$ is the so-called spin parameter. Eq. 2 is based on the data of Alaways (Alaways 1998; Alaways and Hubbard 2001) and Watts and Ferrer (Watts and Ferrer 1987). The Alaways experiment used a motion capture technique to determine the initial conditions of a pitched baseball and to track the trajectory, from which $C_L$ and $C_D$ could be determined for speeds up to approximately 75 mph and for $0.1<S<0.5$. Watts and Ferrer measured the lift force directly from force probes mounted on a spinning ball in a wind tunnel. They measured $C_L$ only at low speed ($v<37$ mph) but with S in the range 0.4-1.0. The SHS parametrization is a rough bilinear fit to both sets of data. RKA uses the parametrization (Adair 2002)

$$\text{RKA:} \quad C_L = 2C_D\,S[1 + 0.5(v/C_D)dC_D/dv] \tag{3}$$

which follows from the argument that the lift on a spinning ball is related to the difference in drag between two sides of the ball which pass through the air at differ-ent surface speeds due to the rotation, an argument first posed by Sir Isaac Newton. RKA argues that Eq. 3 leads to lift values which are in good agreement with the Watts and Ferrer data at low speed, where $C_D$ is relatively constant and equal to about 0.5. Indeed, under these conditions Eqs. 2-3 imply $C_L$=0.21 or 0.20 for SHS or RKA, respectively, for S=0.2. RKA further argues that extending this prescription to higher speeds leads to predictions in reasonable agreement with observations from the game itself (Adair 2002), such as the maximum break of a pitched curveball. However, it is precisely at high speeds—speeds that are typical of pitched or hit baseballs—that the differences between SHS and RKA are most pronounced. This is demonstrated in Fig. 1, which shows the calculated ratio of lift to weight as a function of speed for ω=1800 rpm. While the two curves agree quite well below 40 mph, they diverge for larger speeds, so that in the vicinity of v=100 mph (S≅0.154), the lift is about three times larger for SHS than for RKA. This will have dramatic implications for the distance traversed by a long fly ball.

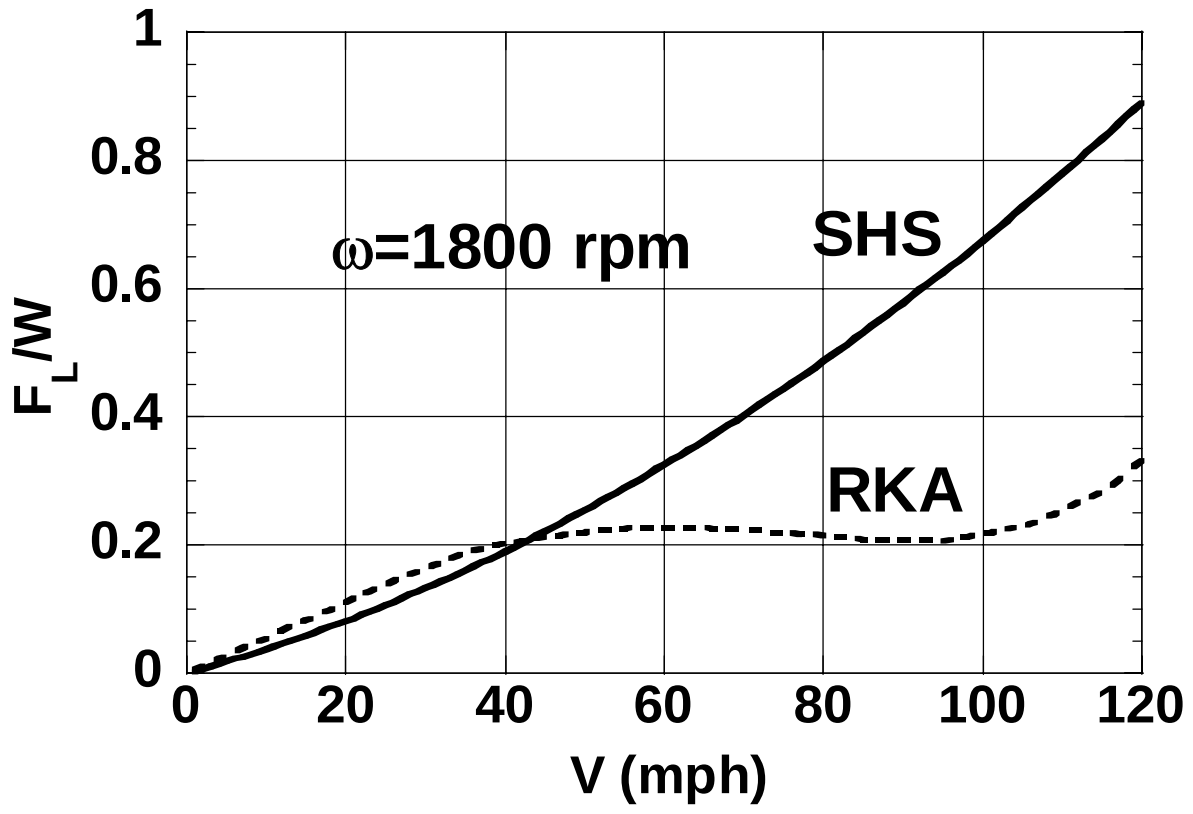

**Fig. 1.** Calculated ratio of lift to weight for ω=1800 rpm.

## 3 Description of the Experiment

The essential technique is to project the ball approximately horizontally with backspin or topspin and to use a motion-capture system to measure the initial velocity and angular velocity and track the trajectory over approximately five meters of flight. Under such conditions, the vertical motion is particularly sensitive to the lift force, which leads to a downward acceleration smaller or larger than *g* when the ball has backspin or topspin, respectively. Additional information is obtained by measuring the total distance traversed by the baseball before hitting the floor, which is approximately 1.5 m above the initial height. The projection device was an ATEC two-wheel pitching machine, in which the speed of the ball is determined by the average tangential speed of the two wheels and the spin is determined by the difference in the two tangential speeds. In the present experiment, we investigated speeds in the range V=50-110 mph and spin rates ω=1500-4500 rpm, corresponding to Reynold's numbers Re=(1.1-2.4) x $10^5$ and S=0.1-0.6. A total of 22 pitches were analyzed, all in the "two-seam" orientation.

## 3.1 Motion Capture Technique

To measure the initial velocity and angular velocity of the ball, a motion analysis system was used. The system, manufactured by Motion Analysis Corporation, consisted of 10 Eagle-4 cameras operating at 700 frames per second and 1/2000 second shutter speed and the EVaRT4.0 reconstruction software. Each camera shines red LED light onto the ball, attached to which is circular dot of retro-reflective tape. The tape reflects the light back at the cameras, which then record the coordinates of the dot in the CCD array. The reconstruction software determines the spatial coordinates in a global coordinate system by using triangulation among the 10 cameras. The cameras were positioned at various heights along a line approximately parallel to and six meters from the line-of-flight of the ball. In order to accomplish the triangulation, the precise position and lens distortions of each camera must be determined from an elaborate calibration scheme. The global coordinate system is defined by positioning an L-shaped rod in the viewing volume of the 10 cameras simultaneously. The rod has four reflective dots located at precise relative locations to each other. With these distances, the software determines a first approximation to the location of each camera. These distances are further refined and the lens distortions determined by waving a wand throughout the tracking volume of the cameras. The wand has three reflective dots at known relative distances. Although the particular calibration software is proprietary, it almost surely uses some variation of the direct linear transformation technique described by Abdel-Aziz and Karara (Abdel-Aziz and Karara 1971) and others (Heikkila and Silven 1997). A typical root-mean-square (rms) precision for tracking a single dot is 1.3 mm. Additional calibrations and consistency checks were performed. A plumb line was used to establish that the *y*-axis of global coordinate system made an angle of $0.16^0$ with the vertical. The clock of the motion capture system was checked against a precisely calibrated strobe light and found to be consistent within 0.5%. A non-spinning baseball was tossed lightly in the tracking volume and the vertical acceleration was measured to be *g* to within 1.5%.

The setup for this experiment is similar in many respects to that used in the pioneering experiment of Alaways (Alaways 1998) but different is some key respects. One difference concerns the deployment of cameras. Alaways used two sets of motion capture cameras. One set tracked the ball over the first 1.2 m of flight and was used to establish the initial conditions; a second set tracked the center-of-mass trajectory over the last 4 m of flight, starting approximately 13 m downstream from the initial position. This allowed a very long "lever arm" over which to determine the acceleration but a short lever arm for determining the initial spin. In the present setup, we used only one set of cameras distributed spatially so to track over the largest distance possible, approximately 5 m. The same set of cameras determined both the initial conditions and the acceleration. Tracking over a larger distance is useful for measuring the spin at small S, since the angle through which the ball rotates over distance D is proportional to SD. For all of the pitches analyzed in the current experiment, the ball completed at least one complete revolution over the tracking region. We gained some redundant information by measuring the total distance R traversed by the baseball while falling through a height of about 1.5 m. The second difference concerns the deployment of reflective markers. Alaways utilized 4 dots on the ball with precisely known relative positions. This allowed him to determine both the initial orientation of the ball and the direction of the spin axis. We experienced great difficulty in tracking more than one dot and therefore used only a single dot, offset from the spin axis by approximately $15^0$. We were not able to measure the spin axis but simply *assumed* it was constrained by the pitching machine to lie in the horizontal plane, perpendicular to the direction of motion. New measurements are planned for early 2006 using multiple dots and those results will be reported at the conference.

## 3.2 Data Reduction and Analysis

A plot of y(t) and z(t) for one of the pitches is shown in Fig. 2, where y and z are the vertical and horizontal coordinates, respectively, of the reflective marker. Standard nonlinear least-squares fitting algorithms were used to fit these data to functions of the form

$$y(t) = y_{cm}(t) + A\sin(\omega t+\phi) \quad (4)$$

$$z(t) = z_{cm}(t) \pm A\cos(\omega t+\phi) \quad (5)$$

where the first term on the righthand side is the center of mass coordinate of the ball and the second term is the location of the rotating dot with respect to the center of mass. The center of mass motion can be calculated numerically given the initial values of the $y_{cm}$ and $z_{cm}$ positions and velocities and the lift and drag coefficients. The rotation of the dot is characterized as an oscillation with amplitude A, angular frequency $\omega$, and initial phase $\phi$, with the +/– sign applying for backspin/topspin. Therefore, our fitting procedure was used to determine nine parameters:

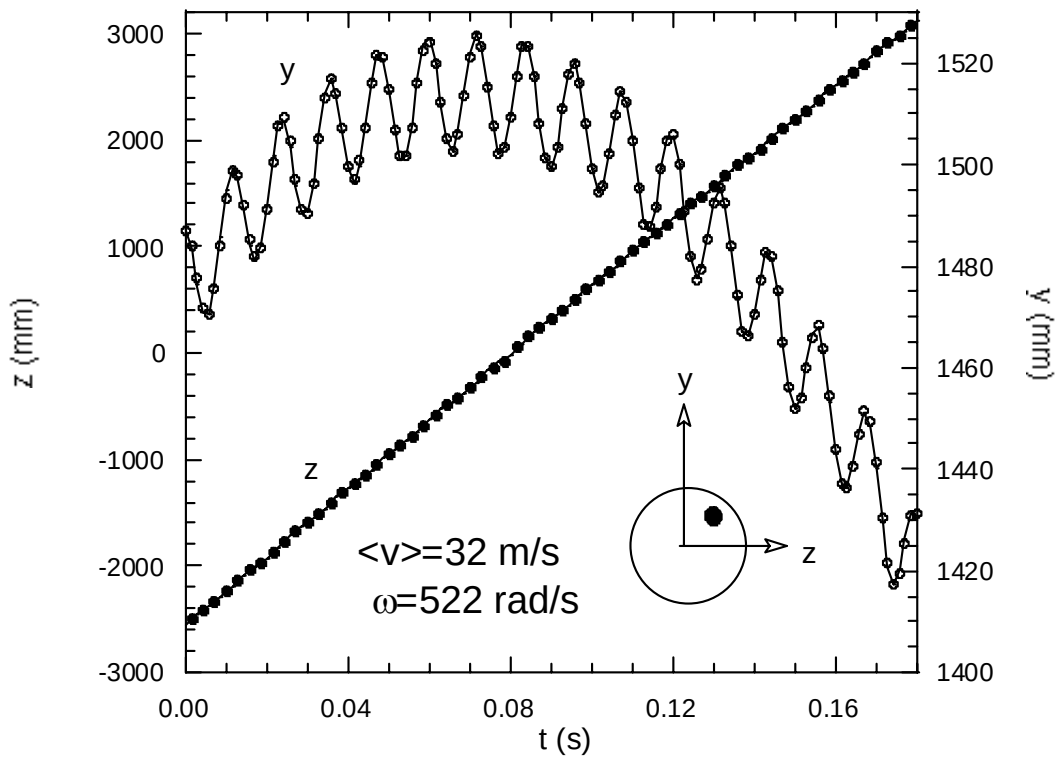

**Fig. 2.** Trajectory data for one of the pitches projected, where y and z are the coordinates of the dot on the ball in the coordinate system shown in the inset. For clarity, the plot for the z coordinate displays every other point. The ball is projected at a slight upward angle to the +z direction with topspin. Curves are least-squares fits to the data using Eqs. 4-5.

the four initial values, the three rotation parameters, and $C_L$ and $C_D$, both of were assumed to be constant over the 5 m flight path. Depending on the initial speed, each fit had 100-300 degrees of freedom, including the range R. Our analysis procedure is quite similar to that used by Alaways. The curves in Fig. 2 are the results of the fit to the data. The rms deviation of the fit from the data in Fig. 2 is 0.4 mm for y(t) and 12.7 mm for z(t), which values are typical of the fits for the other pitches. The inferred values of $C_L$ and $C_D$, which represent the main results of our experiment, are presented in Figs. 3-4. The error bars on the values are estimates based on the rms deviation of the data from the fit and the calibration tests.

# 4 Results and Discussion

## 4.1 Results for $C_L$

Our results for $C_L$ are shown in the top plot of Fig. 3. Also shown are the SHS parametrizaton and the earlier results upon which the SHS parametrization is based,

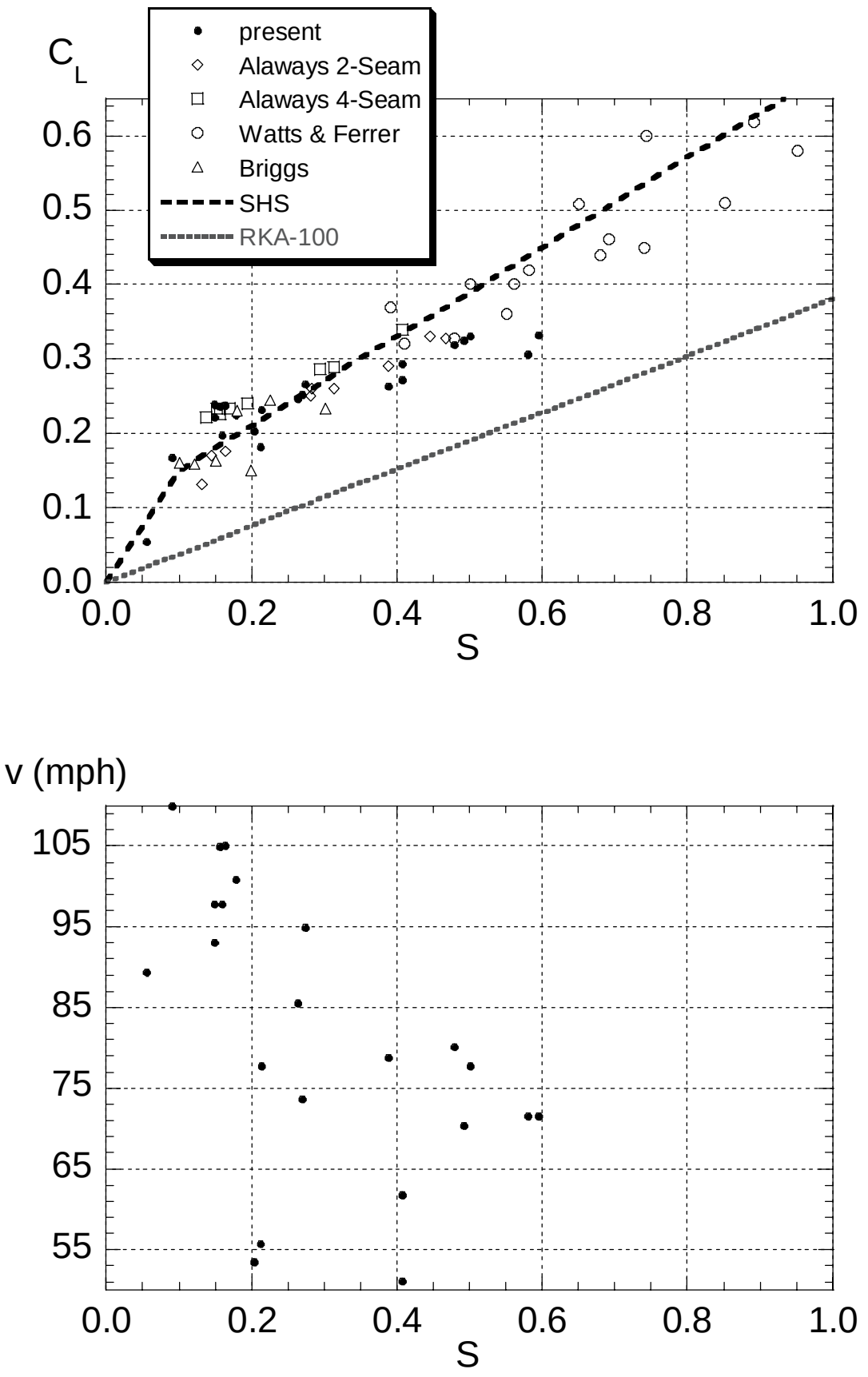

**Fig. 3.** Results for $C_L$. The top plot show $C_L$ from the present and previous experiments, along with the parameterizations of SHS and RKA, the latter calculated for a speed of 100 mph. Bottom plot shows v vs. S for the present data.

including the data of Alaways, Watts and Ferrer, and Briggs (Briggs 1959), as corrected by Hubbard (Sawicki et al. 2005). In the region S<0.3, which is the region most relevant for long fly balls, our data are in excellent agreement with SHS and the earlier data. In fact, only for S>0.5 do our results start to deviate from SHS, which is constrained primarily by the high-S data of Watts and Ferrer, most of which were taken at speeds below 25 mph. Interestingly, the present data near S=0.5 agree with

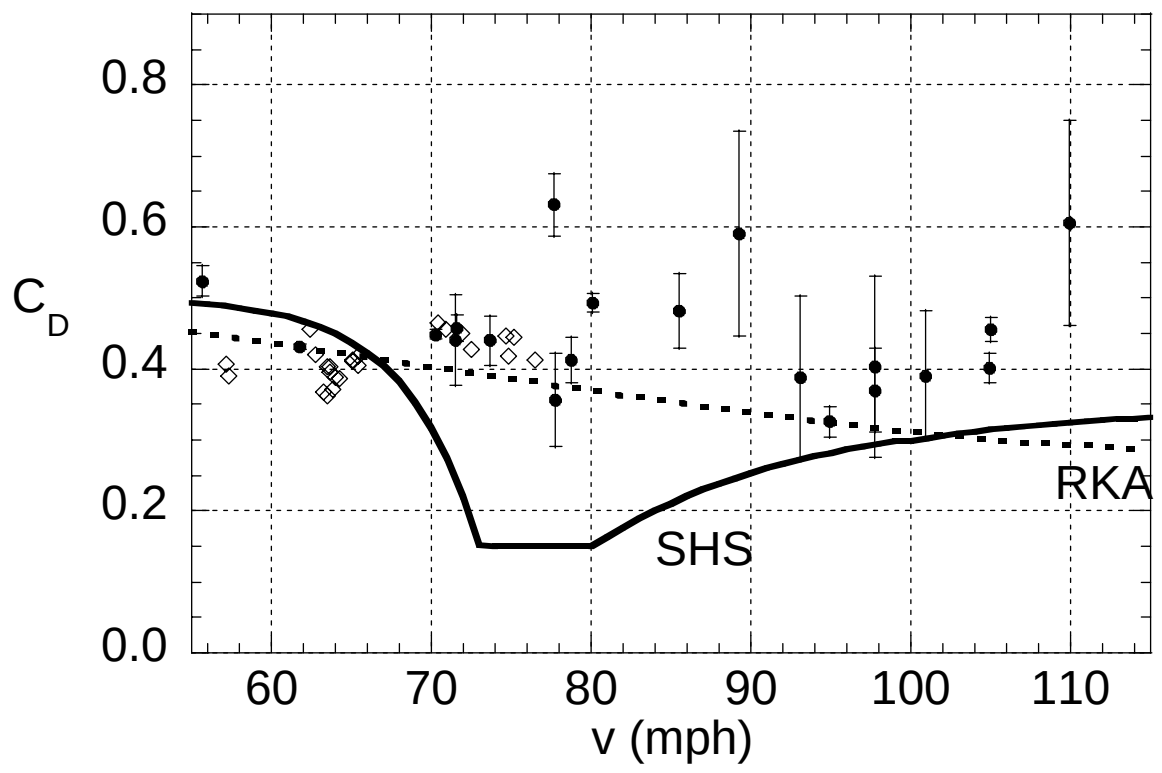

**Fig. 4.** Results for $C_D$ from the present (closed points) and Alaways' (open points) motion capture experiments, along with the parameterizations of SHS and RKA.

Alaways and are below the general trend of the Watts and Ferrer data. Perhaps this is an indication of a dependence of $C_L$ on v, for a fixed S. On the other hand, the bottom plot of Fig. 3 shows the correspondence between v and S for the present experiment. In the range S=0.15-0.30, the values of $C_L$ are tightly clustered around 0.20, despite a variation in v between 50 and 110 mph. This seems to suggest only a weak dependence of $C_L$ on v, at least in that regime. Also shown in Fig. 3 is the parametrization of RKA for a speed of 100 mph, a speed typical of a well-hit ball. In the vicinity of S=0.17, corresponding to a backspin of 2000 rpm, the RKA curve falls well below the data. We conclude that the present data support SHS and do not support RKA.

## 4.2 Results for $C_D$

Our results for $C_L$ are shown in Fig. 4. Also shown are the SHS and RKA (Adair 2002) parametrizations and the motion capture data of Alaways (Alaways 1998). There is considerably more scatter in our results for $C_D$ than for $C_L$, reflecting the generally larger rms deviation of the fit for z than for y. It is easy to see from Fig. 2 that our technique is much more sensitive to $C_L$, which is determined by the curvature of y(t), than to $C_D$, which is determined by the curvature of z(t), the latter nearly completely masked by the large linear term. Therefore any comments we make about $C_D$ based on the present data must be considered speculative. That said, our results are not in agreement with SHS but they in rough agreement with RKA. Interestingly, our results agree very well with the Alaways motion capture data in the regime where they overlap below 80 mph, and neither observed a large dip in the region of 70-90 mph, the so-called drag crisis. The dip comes from an analysis of high-speed video of pitched baseballs from the Atlanta Olympics (Alaways, Mish, and Hubbard 2001). On the other hand, the motion capture data in the 70-80 mph range generally were taken with considerably larger spins (4000 rpm) than found with pitched baseballs. This supports the notion of spin-dependent drag (Adair 2002) which might smooth out the drag crisis, giving a curve more similar to RKA than to SHS.

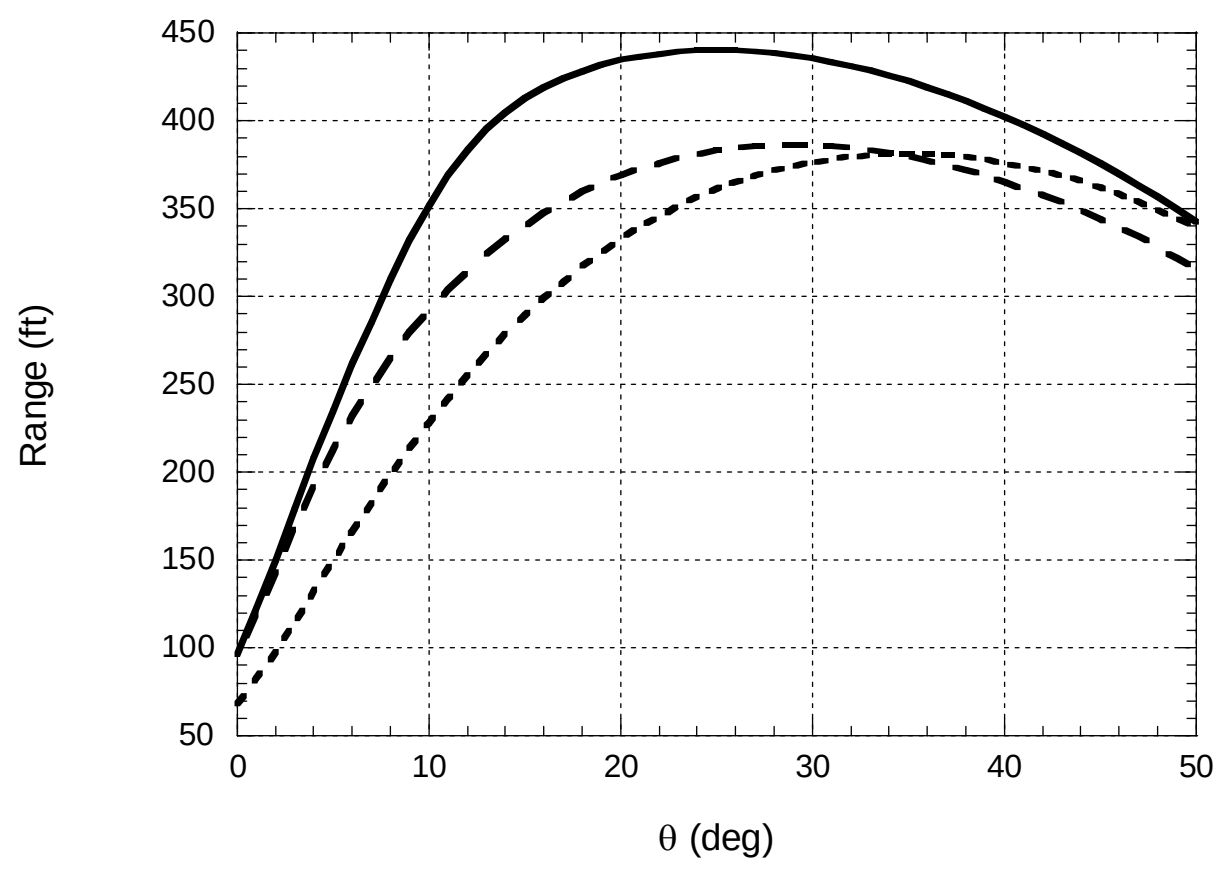

**Fig. 5.** Range of a hit baseball as a function of initial angle, with an initial speed of 100 mph, backspin of 1800 rpm, and height of 3 ft. The three curves utilize different prescriptions for lift and drag, as discussed in the text.

### 4.3 Implications for the Flight of a Baseball

Finally we explore briefly the implications of our results for the flight of a baseball. In Fig. 5 a calculation is shown of the maximum range for a hit baseball as a function of the takeoff angle for three different parameterizations of lift and drag. The solid curve uses SHS for both lift and drag; the short-dashed curve uses RKA for both lift and drag; and the long-dashed curve uses RKA for drag and SHS for lift. The latter prescription is the one most consistent with the present data.

## 5 Summary and Conclusions

We have performed an experiment utilizing high-speed motion capture to determine the effect of spin on the trajectory of a baseball. From these data, we determine with good precision values for the $C_L$ over the range $50<v<100$ mph and $0.1<S<0.6$. Our $C_L$ is in excellent agreement with the SHS parameterization for $S<0.3$ but falls below SHS for larger S. Our results are not consistent with the RKA parameterization. With considerably less precision, we have also extracted values for $C_D$, which do not support a sharp drag crisis in the 70-90 mph range.